\documentclass[breakaddress,preprint,twocolumn,reprint]{revtex4-1}

\usepackage[pdftex]{graphicx}
\usepackage{dcolumn}
\usepackage{bm}

\usepackage{empheq}
\begin{document}

\title{Electrically charged droplet: Influence of the generator configuration and the nature of the liquid}

\author{Martin Brandenbourger}
\author{St\'ephane Dorbolo$^\dag$}
\address{GRASP, Physics Department, University of Li\`ege, B-4000 Li\`ege, Belgium.}

\address{$^\dag$F.R.S.-FNRS, GRASP, Physics Department, University of Li\`ege, B-4000 Li\`ege, Belgium.}

\begin{abstract}
We studied how to charge droplets by induction and how to maximize this charge. In order to aim this objective, we developed an innovative device that avoids non-linear effects and that is able to charge liquids of different nature. The device developed resembles a planar capacitor.  The influence of the nature of the liquid (i.e. presence of ions in solution, polarity, surface tension and conductivity) on the charge induced was measured. We deduced that there exists a threshold of electrical conductivity for the fluid below which it does not charge according to the ``perfect conductor" model.\\
\end{abstract}

\maketitle

\section{Introduction}

Basically, rain falling from thunderclouds is composed of electrically charged droplets \cite{Saunders}. Since Millikan's experiment \cite{millikan}, which permits to measure the ratio between charge and mass of an electron via the study of charged droplets, the electrically charged droplet were intensively studied regarding applications. Indeed, they are the basis of many applications such as electrospray \cite{spectromass}, electrowetting \cite{electrowetting}, or electrospinning \cite{electrospinning}. Electrically charged droplets have also been studied from a fundamental point of view. Indeed, electrically charged droplets lead to impressive behaviors such as the Coulomb explosion \cite{Coulinstab} or particular interactions between droplets \cite{coalesc}. \\

Three different ways to charge droplets can be found in literature : triboelectricity, radioactive radiations or charge by influence. The first two manners are frequently used in Millikan experiments. They have the advantage to charge droplets easily, without any ``complicated"  set-up. However, both techniques do not allow to predict the charge induced in the droplet. On the contrary, the charge by influence allows to determine precisely the charge in the droplet. Indeed, as we will show, the geometry of the set-up (and the voltage applied to it) correspond to a given amount of charges induced in the droplet. However, the mechanism of charging by induction a liquid is more complex to explain than the charge by induction in a solid conductor. Indeed, one of the most important issue is that the nature of the charges induced in the droplet has not already been clearly identified.\\

In this paper, we explain how to develop an innovative apparatus which permits to charge droplets by influence. This device allows to induce a given charge in a droplet that has radius between $1$ mm and $3$ mm. The geometry of the apparatus has also been chosen to avoid any non-linear phenomenon. Indeed, this configuration allows to describe the charge induced in the droplet with a simple law. Moreover, we will see that this configuration allows to charge droplets made of different liquids whatever their surface tension and their wetting properties. \\

After a description of the experimental set-up, we present the study of the geometry of our new apparatus and the influence of the chemical composition of the liquid used for droplets. Both objectives allow to determine how to maximize the charge induced in the droplet and to understand in a better way the mechanism to charge liquid by induction.\\
\section{Experimental set up}

\subsection{Charged droplet generator} \label{generator}

As stated above, the charged droplet generator is based on the charge by induction mechanism. This kind of device generates an electric field which attracts positive (or negative) charges in a droplet or/and extracts negative (or positive) charges out of a droplet (see, for example \cite{classicgenerator}). \\

\begin{figure}[htbp]
\includegraphics[scale = 0.22]{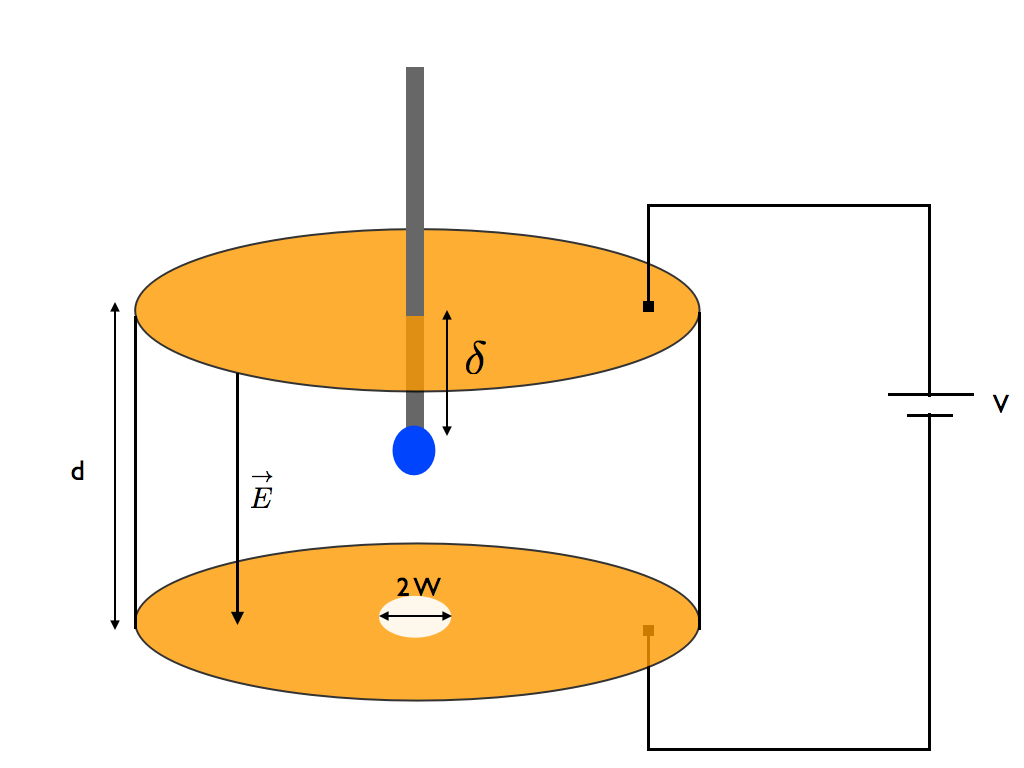}
\caption{(Color online) Sketch of the charged droplet generator. Both metal disks are represented by  yellow circles. They are separated by a polycarbonate ring (not represented). The needle is represented by the grey vertical line. The voltage between both plates allows to generate an electric field ($\stackrel{\rightarrow}{E}$) and then, to charge by influence the pendant droplet. Note that $\delta$, $d$ and $W$ are parameters to optimize.}
\label{dropgenerator}
\end{figure}

The geometry of our charged droplet generator was chosen to answer to several requirements. On one hand, its geometry should avoid nonlinear phenomenon as much as possible. But on the other hand, it had to be usable for various liquids, which implies (i) to be able to generate high electric fields (ii) to be able to deal with low surface tension fluids. In order to tackle these constraints, the charged droplet generator was made of two metal disks, parallel and positioned along a vertical axis (see Fig. \ref{dropgenerator}). The top plate was pierced by a needle from which droplets was formed. A hole was drilled in the center of the bottom plate to allow the droplet to fall away from the charged droplet generator. When a given voltage is applied, electrodes generate a given electric field which induces a given charge in the pendant droplet attached to the needle.\\

Four parameters allow to describe the geometry: the piercing distance $\delta$ of the needle into the top plate,  the radius $W$ of the hole in the bottom plate,  the distance $d$ between both plates, and the voltage $V$ between the circular plates. All these parameters have been studied to optimize the charge induced in the droplet. A compromise had to be found to have large electric field but keeping a linear dependence of the charge with the applied voltage.

\subsection{Charge detection set-up : The Faraday cup}

\begin{figure}[htbp]
\includegraphics[scale = 0.23]{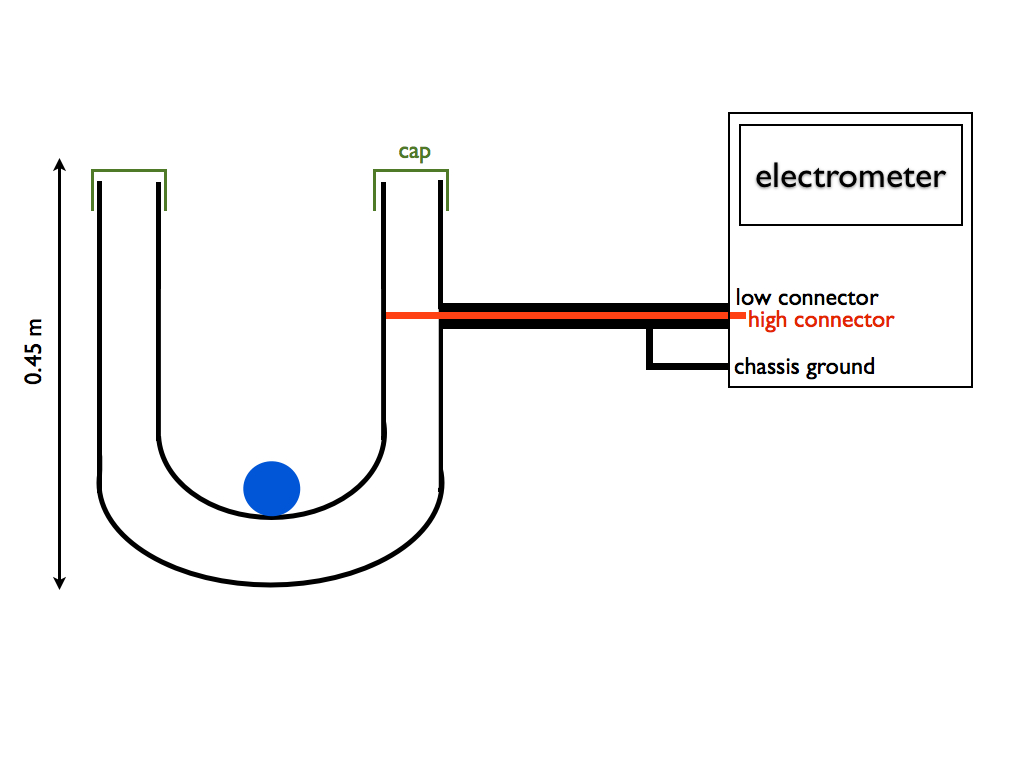}
\caption{(Color online) Illustration of the Faraday cup. The droplet falls into the inside of the cup and discharges in the inner cup. The electrometer measures the charge of the droplet in comparison to the chassis ground.}
\label{faradaycupstat}
\end{figure}

The measurement of the charge of the droplet is essential. This was performed by using a Faraday cup. This kind of device is usually used in ions beam experiments. Contrary to this scientific field, the Faraday cup developed here aims to measure the charge of mesoscopic objects, i.e. charged droplets. The Faraday cup created to achieve this goal is schematized in Fig. \ref{faradaycupstat}. It was composed of two concentric cups. The space between both cups was covered by a cap to protect the inside of the device from external charges. The inside cup was connected to the input ``high" connector of an electrometer (Keithley 610C) while the outside cup was connected to the input ``low" connector and to the chassis ground. The electrometer was set in coulombmeter mode. When a droplet falls into the Faraday cup, it discharges on the inside of the cup.\\ 

\section{Influence of the generator configuration}

This section aims to study the influence of the geometry ($d$, $W$, $\delta$) of the charged droplet generator on the charge induced in the droplet. All experiments performed for this purpose was made with bidistillated water.\\

\subsection{Theoretical justification of the chosen geometry} \label{essence}

Charged droplets generators have already been studied in the past years. For example, Jones $et$ $al$ have proposed a geometry for a charged droplet generator \cite{classicgenerator} using charge by induction. Their geometry, composed of one needle and one plate, is commonly used in electrospray devices \cite{electrospraydevice}. In the context of this kind of device, they proposed also a theoretical approach to predict the charge induced in the droplet. Assuming that the generator and the liquid are perfect conductors, it is possible to calculate the electric field induced in the charged droplet generator. Knowing the electric field in the droplet, they calculate the charge density in the droplet and thus the total charge in the droplet. This approach will be also followed here. \\

We propose here a geometry which looks like a planar capacitor. Indeed, theoretically, this kind of configuration allows to avoid non-linear effects and to describe the charge induced in a droplet by very simple laws. The device described in the section \ref{generator} is a planar condensation if one considers $\delta$ and $W$ small enough. If it is the case, it is possible to calculate the charge induced in the droplet with the same arguments as Jones $et$ $al$. \\

The charge by unit area on a planar capacitor is :

\begin{equation} \label{eq1}
\frac{Q}{S}=\frac{V \epsilon_0}{d}
\end{equation}
with Q the charge of the droplet, S the surface of the droplet and $\epsilon_0$ the vacuum permittivity. Assuming that the droplet is a perfect conductor, the total charge induced in the droplet is :

\begin{equation} \label{eq2}
Q=\frac{V \epsilon_0 4 \pi r_d^2}{d} \text{}
\end{equation} 
with $r_d$ the droplet radius.  This later equation is a simple relation which allows to easily predict the charge induced in a droplet.  By assuming that our device corresponds to this geometry, we will firstly study the influence of $d$ and $V$ on the device and compare the experimental behavior with \ref{eq2}.\\

Unfortunately, in practice, a charged droplet cannot be created from a planar capacitor. Indeed, a droplet formed on the top plate wet the surface before falling from the top plate and the bottom plate needs a hole in its center to allow the droplet to fall away from the generator. That is why the device developed in this study is composed of a needle which pierces the top plate and a hole in the bottom plate. In consequence, we also have study the influence of $W$ and $\delta$ on the validity of the hypothesis.\\

\subsection{Influence of $d$ and $V$}

\begin{figure}[htbp]
(a)
\includegraphics[scale = 0.6]{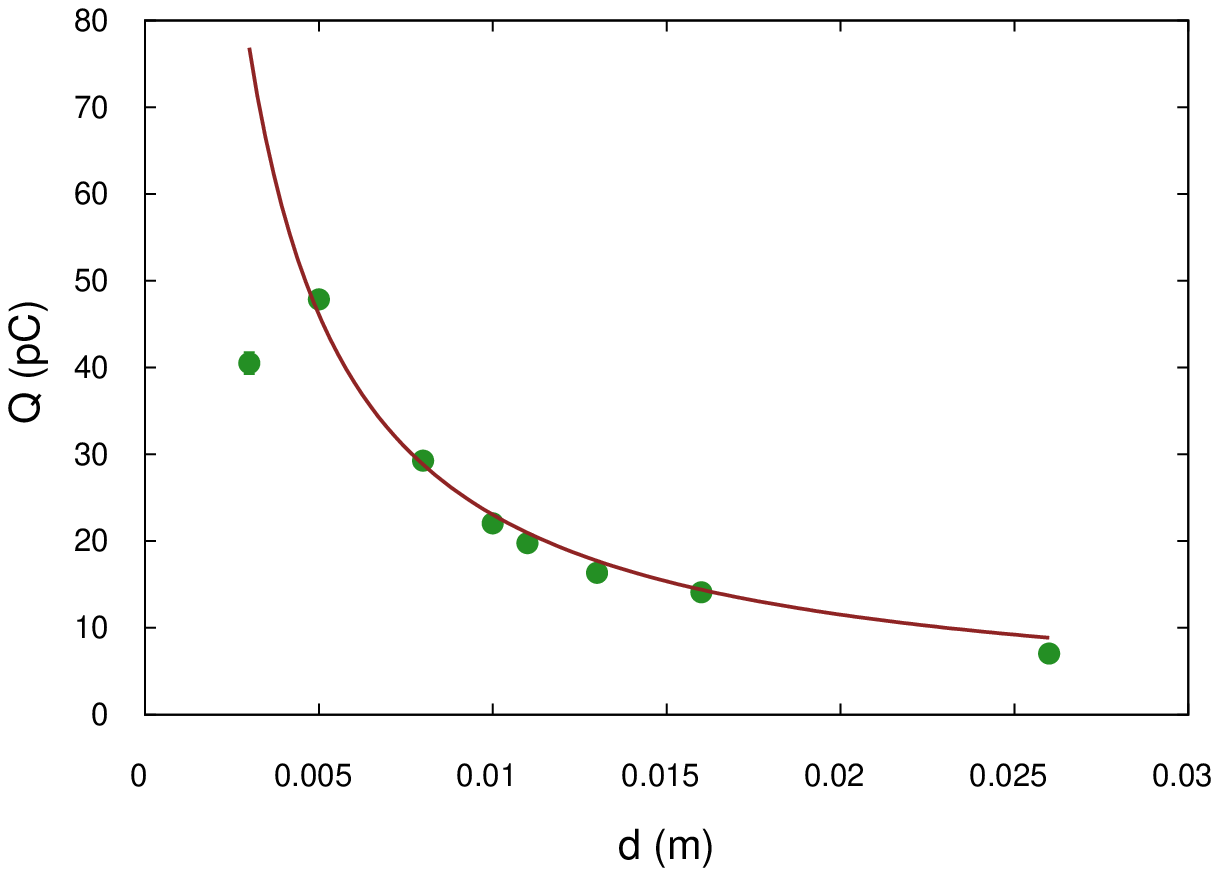} \\
(b)
\includegraphics[scale = 0.6]{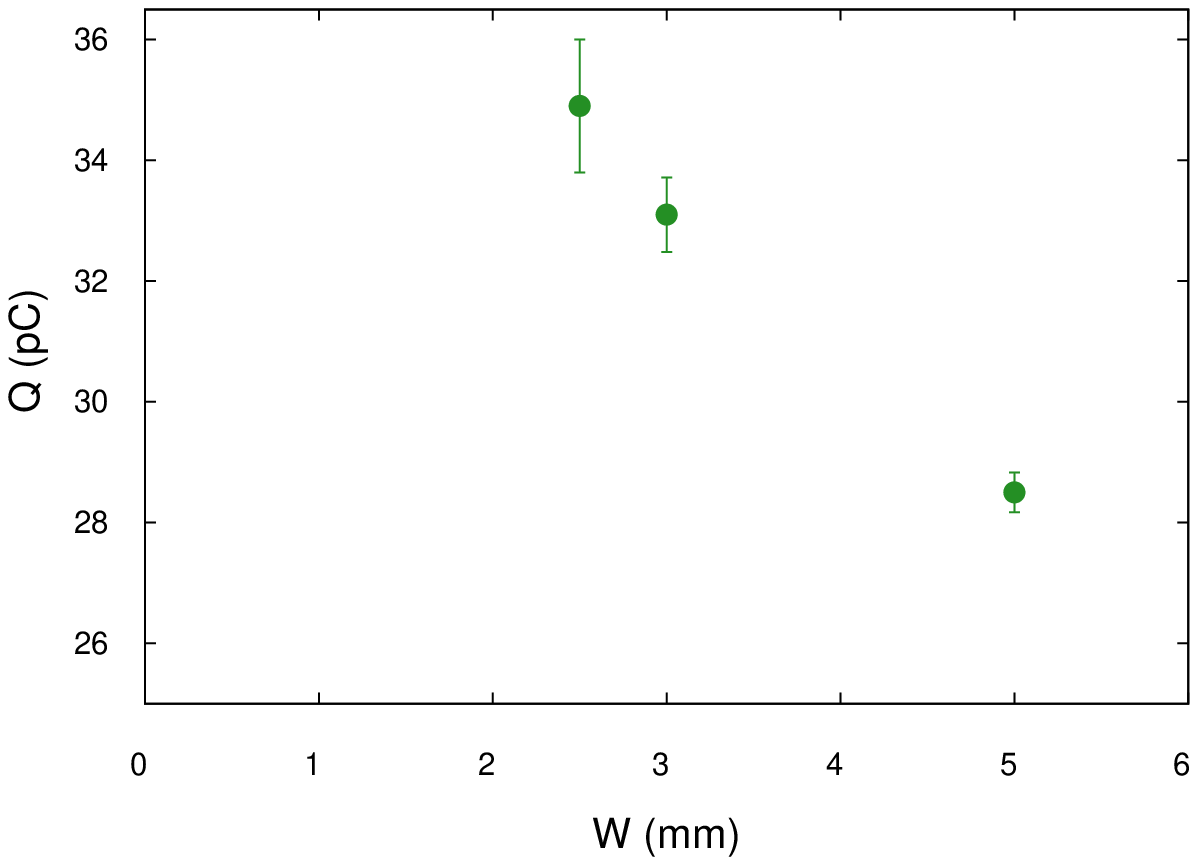} 
\caption{(Color online) (a) Charge induced in the droplet as a function of the distance between both plates $d$.  Error bars are hidden by the points size. The red curve is a fit between $d=5$ mm and $d=26$ mm. (b) Charge induced in the droplet as a function of the radius of the hole in the bottom plate, $W$.  Measurements indicates that a rise in $W$ tends to reduce the charge induced in the droplet.}
\label{gaphedmoisdelta}
\end{figure}

First of all, the influence of $d$, the distance between both disks, is considered. Figure \ref{gaphedmoisdelta} (a) presents the charge measured on the droplet as a function of $d$. The red curve is the fit made on the data between $d=5$ mm and $d=26$ mm. In this case, in regard to the errors bars, the fit matches with the measurements. However, the first measurement, at $d=3$ mm, does not match with the fit. Consequently, it seems that the law is in great agreement with the measurements except for small distance $d$. This behavior can be explained by the study of $\delta$ and $W$ as we will see later.\\

Afterwards, the variation of the voltage, $V$, between the metal disks has been studied. Figure \ref{difliq} shows the linear behavior of the charge in the droplet as a function of the voltage. The measurements made on water are plotted in red triangles.  Measurements confirm the linear behavior stated by the Eq. (\ref{eq2}) in the range of voltage studied (i.e. from 0 to 200 V).\\

Finally, the variation of the polarization between both metal disks has been studied. As predicted in Eq. (\ref{eq2}), this variation only induce a change in the sign of the charge. Indeed, when the higher voltage is on the top disk, the droplet is charged positively and when the higher voltage is on the bottom disk, the droplet is charged negatively.\\

\subsection{influence of $W$ and $\delta$ }

We now consider the case of small distance $d$ between the metal disks. In this case the presence of the needle (i.e $\delta$) becomes significant. Consequently, the geometry becomes different from the case of a planar capacitor. Indeed, the system approach the geometry described by Jones $et$ $al$ (only a needle and a plate) and for very small $d$, the droplet begins to be out of the charged droplet generator before it detaches from the needle (which implies that the electric field does not apply on all of the droplet). Moreover, for short distances between circular plates, the hole in the bottom plate is not anymore negligible. As a consequence, the electric field generated by the system is smaller than expected. These arguments are confirmed by the measurement at very small $d$ (see Fig.  \ref{gaphedmoisdelta}) and by the study on the effect of the variation of $W$.  \\

The effect of $\delta$ is directly linked to the distance $d$. Indeed, $\delta$ begins to be non-negligible  when it is comparable to $d$. Practically, the geometry of our charged droplet generator does not allow to vary the distance $\delta$. Indeed, the syringe has to be weld to the top disks to allow a good electrical connection. During the measurements on the effect of $d$, $\delta$ was fixed to $3$ mm.  The study of the charge induced as a function of $d$ indicates that the measurements differs from the behavior described by the Eq. (\ref{eq2}) when $d<5$ mm. As a consequence, the assumption that supposes that $\delta$ begins to be non-negligible when it is comparable to $d$ is confirmed.  \\

The effect of $W$ is shown in Fig. \ref{gaphedmoisdelta} (b). Measurements was made for $d=10$ mm and $\delta=5$ mm. The figure shows that an increase in $W$ is directly linked to a decrease of the charge induced of the droplet. In consequence, it indicates that the hole in the bottom plate affects the geometry assumed in section \ref{essence} as soon as $W$ is to large. Note that the radius of the hole cannot be too small. Indeed, the droplet have to be able to leave the generator. Consequently, $W$ has to be larger than the radius of the droplet.\\  

\subsection{Geometrical optimization of the charged droplet generator}

According to Eq. (\ref{eq2}), it seems that, for a given droplet radius, $r_d$, the optimal configuration is reached when $d$ is about zero and $V$ is as important as allowed by the power supply. In practice, the charge induced in the droplet begins to decrease for small $d$ in compare to $\delta$. The optimal $d$ is then defined as the limit where the Eq. (\ref{eq2}) remains valid. In our configuration, this limit is $d=5$ mm. Once $d$ has been minimized, it is also possible to maximize the charge by rising V. This optimization is also limited. Indeed, for a too large voltage, a Taylor cone is formed, and droplets are ejected from the needle before reaching their maximum size. This limit correspond to the formation of an electrospray. It is expressed as follow :

\begin{equation} \label{eq3}
Q_R=2 \pi \sqrt{16 \gamma \epsilon_0 r_d^3}
\end{equation}
with $\gamma$ the surface tension of the liquid. Eq. (\ref{eq2}), indicates also that an increase in $r_d$ increase the charge induced in the droplet. However, this parameter is generally fixed by experiment requirements. Finally, $W$, the radius of the hole in the bottom plate has to be reduced to maximize the charge induced in the droplet. However, it has to be large enough to allow the droplet to fall away from the droplet generator. Consequently, its value has to be calculated to be as short as possible in regard to the droplet radius. In consequence, our optimal set of parameters is $d=5$ mm, $\delta=2$ mm and $W=2.5$ mm.\\

To sum up, the measurements demonstrate that the behavior of the charged droplet generator is in agreement with the ``perfect conductor" hypothesis and, in a certain range, with the planar capacitor hypothesis. However, the bidistillated water is not usually described as a perfect conductor. In order to understand this apparent paradox, we were interested in the influence of the nature of the liquid on the charge induced in the droplet.\\

\section{Influence of the liquid}

\begin{figure}[t]
\includegraphics[scale = 0.65]{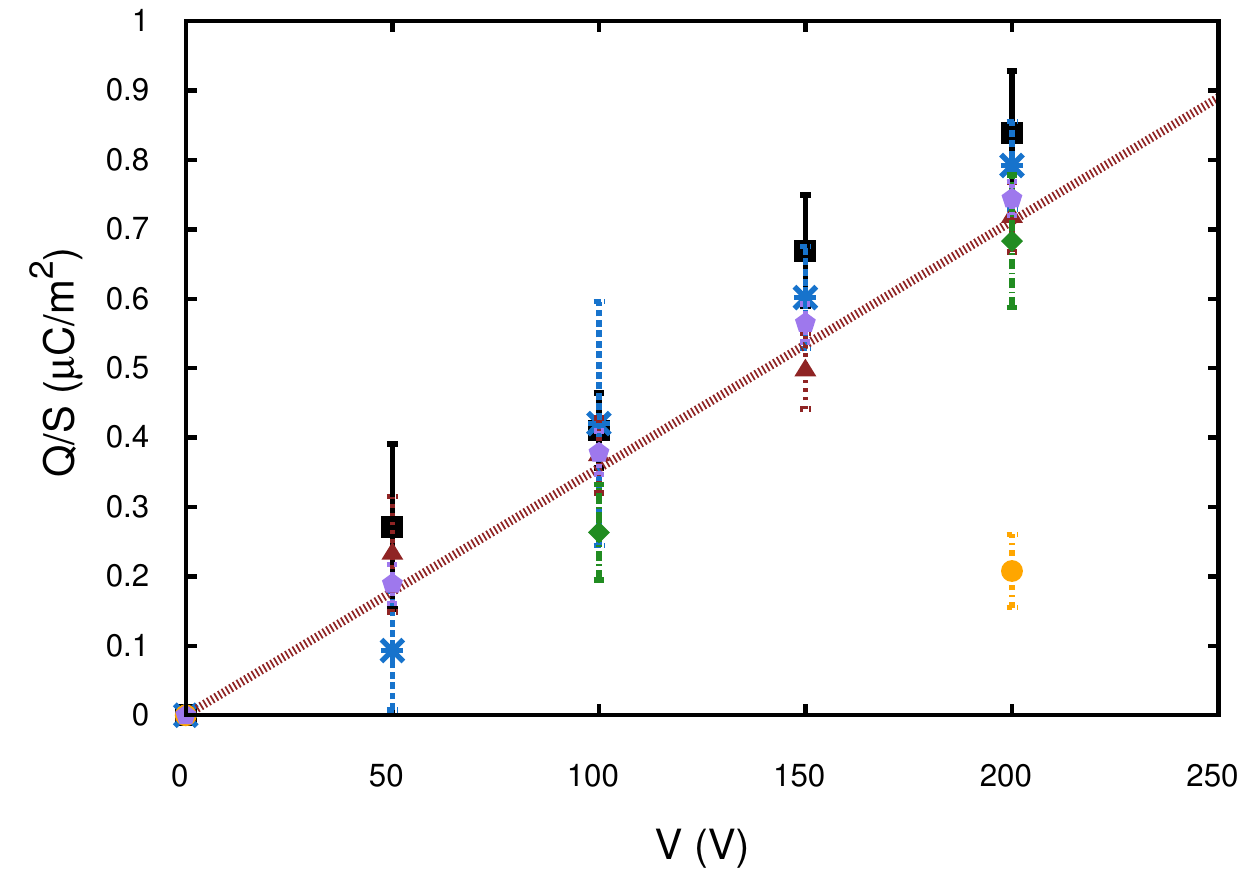}
\caption{(Color online) Charge induced in a droplet, normalized by its radius, in function of the voltage applied on the charged droplet generator. Red triangles, purple pentagons, Black squares, Blue stars, Green diamonds and Yellow circles represent respectively measurements made on bidistillated water, Triton-x-100, acetone, ethanol, octane and silicon oil. The red dotted line is a fit on these measurements regarding to the error bars.}
\label{difliq}
\end{figure}

\begin{table*}[ht]
 \begin{center}   {
\begin{tabular}{|c|c|c|c|c|c|c|c|c|}

  \hline
 &  bidistilled water & HCl (1M) & Triton X-100 &  Acetone &  Ethanol  &  Octane  &  Silicone oil  \\    \hline
 $\sigma$ (mS/cm) &  $5.5$ $10^{-5}$  & 330 & $5.5$ $10^{-5}$ & 21  &  11 &  $10^{-6}$  & $10^{-11}$ \\   \hline
  $\gamma$ (mN/m) & $70.8$ & 63.6 & $32.0$ &  $23.7$ & $21.7$  &  $21.8$  & $19.0$ \\
  \hline
    Polarity & polar & polar & polar & polar  & polar &  non-polar  & non-polar  \\
  \hline
   Ions in solution & yes & yes & yes & no  & no  &  no  & no\\
  \hline
\end{tabular}}
\end{center}
\caption{\label{tab1} Chart of the conductivity ($\sigma$), the surface tension ($\gamma$) , the polarity and the presence (or not) of ions in solution of the liquids charged. All these measurements as been made at 20 Ò\ensuremath{^\circ}Ó C. The conductivity values are drawn from \cite{conduc} and \cite{conduc2} while the surface tension measurements has been made by pendant droplet measurements on a CAM 200 (KSV Ltd). \\}
\end{table*}

In order to check the domain of validity of the ``perfect conductor" hypothesis, several liquids was used to obtain charged droplets. The fluids tested were : bidistillated water, a mixture of water and Triton X-100, ethanol, acetone, octane and silicone oil (with a viscosity of 1.5 cSt, near the water viscosity). Each of these liquids have different fluid and electrical properties. In Table \ref{tab1}, the conductivity $\sigma$, the surface tension $\gamma$, the polarity and the presence of ions in solution are reported. \\

The Fig. \ref{difliq} shows results of measurements on the charge induced in a droplet for a voltage applied on the charged droplet generator between 0 and 200V. The charge of the droplet was normalized by the surface of the droplet. This normalization is explained by the fact that the different liquids do not have the same surface tension. Accordingly, droplets generated by the charged droplet generator do not have the same size. Then, measurements have to be normalized by the surface of the droplet to be comparable, as stated by Eq. (\ref{eq2}). The volume of one droplet was deduced by measuring the volume accumulated by 100 droplets.\\

The first measurements (i.e. measurements which aim to study the geometry of the charged droplet generator) was made with bidistillated water and are in agreement with the ``perfect conductor" hypothesis. Moreover, measurements made with a solution of HCl (1M) lead to the same amount of charges induced in a droplet. The amount of charges induced in a water droplet is thus the same than the case of a good conductor. Besides, bidistillated water has specifics properties. It is a polar liquid with ions in solution and a high surface tension. Consequently, other liquids was used to charge droplets in order to check the influence of these properties.\\

The liquid named Triton X-100 in the Table \ref{tab1} is bidistillated water with an addition of 1CMC (critical micelle concentration, which correspond to 0.23mM) of the non-ionic surfactant Triton X-100. This liquid was used to check the influence of non-ionic surfactant on the charge induced in the droplet. Ethanol and acetone do not possess ions in solutions, it is then possible to check whether the charge in the droplet is due to ion migration. Both liquids are also polar. On the contrary, octane does not possess ions in solution and is apolar. It is thus possible to verify the influence of the molecules polarity. Finally, silicone oil at 1.5 cst is a liquid with viscosity near water but with a smaller electrical conductivity than any other liquids tested.\\

The Fig. \ref{difliq} shows that almost all of the liquids tested become electrically charged in the same way. Because of the errors bars, no distinction can be made between the charge induced in the different liquids.The silicone oil is the only tested liquid which does not charge as well as the others. Indeed, it does not charge significantly below 200V. Before going in further details, it must be remembered that these observations are made between 0 and 200V, and thus at a small voltage compare to the electrospray limit \cite{electrospraydevice}. In this configuration, it seems that it is not surfactant, ions in solution or molecules polarity which leads the droplet to charge. A later important parameter in this system is the conductivity of the liquid used. Our measurements allow to determine the limit of electrical conductivity below which the ``perfect conductor" hypothesis is valid. This limit is located between $10^{-6}$ mS/cm and $10^{-11}$ mS/cm.

\section{Conclusion}

In conclusion, an innovative electrically charged droplet generator has been developed. This device was optimized to generate charged droplets from any liquids while avoiding non-linear effects. This study has also underlined how to optimize the different parameters of the generator. It has been proved that the perfect conductor approach is in agreement with the measurements made on the apparatus at a voltage between 0 and 200V. Moreover, a detailed study was performed on the influence of the liquid on the charge induced in the droplet. It results from this study that a large panel of liquids could be charged by the generator. We determine that to be easily charged accordingly to the ``perfect conductor" hypothesis, a liquid has to have a conductivity larger than $10^{-7}$ mS/cm.

\section*{Acknowledgements}
MB is financed by Micromast (AIP (P7/38) BELSPO ). SD is an F.R.S.-FNRS research associate. This study was financially supported by the ESA topical team Poladrop and ``Fonds Sp\'ecifiaux de la Recherche" (ULg).

\end{document}